\documentclass[sigconf]{acmart}

\AtBeginDocument{%
	}

\setcopyright{acmcopyright}
\copyrightyear{2023}
\acmYear{2023}
\acmDOI{XXXXXXX.XXXXXXX}

\acmConference[Parallel and Distributed Computing]{}{Spring 2023}{Hefei, China}
\acmBooktitle{Parallel and Distributed Computing}
\acmPrice{15.00}
\acmISBN{978-1-4503-XXXX-X/23/08}

\usepackage[linesnumbered,ruled]{algorithm2e}
\usepackage{subfig}
\usepackage{overpic} 
\usepackage{url}
\usepackage{float}
\usepackage{caption}
\usepackage{verbatim}
\usepackage{listings}
\begin{document}
	
    \title{Technique: Simulate Bumblebee and Extend It to Support LE Coded PHY in BLE version 5}
	
    \author{Hao Zhao}
	\email{zhaohao99@mail.ustc.edu.cn}
	\affiliation{%
		\institution{University of Science and Technology of China}
		\city{Hefei}
		\state{Anhui}
		\country{China}
	}
	

	
	
	\begin{abstract}
		


        Bumblebee \cite{xu2023bumblebee} is an innovative backscatter system that leverages productive Bluetooth Low Energy (BLE) carriers to generate ZigBee transmissions. It improves upon previous systems like FreeRider and Interscatter \cite{iyer2016inter} by eliminating restrictions on exciters and receivers, resulting in reduced deployment costs and increased throughput. To enable long-distance signal transmission, BLE version 5 introduces LE Coded PHY. However, the paper proposing Bumblebee only evaluates it with BLE version 4, which utilizes LE Uncoded PHY. In this paper, we initially simulate Bumblebee using MATLAB to assess its design. Subsequently, we extend Bumblebee to support LE Coded PHY in BLE version 5 and conduct experiments to verify its performance.
	\end{abstract}


	
	\maketitle
	
	\section{Introduction}
        \label{sec:introduction}


        Ambient backscatter systems have gained significant attention in recent years due to their ability to facilitate communication among billions of Internet of Things (IoT) devices with minimal power consumption. Unlike traditional backscatter systems such as RFID, ambient backscatter systems do not require dedicated carriers and instead utilize ambient signals. Additionally, many IoT devices that support widely used wireless protocols like Zigbee and BLE can also function as backscatter receivers. This approach offers a cost-effective and versatile solution for IoT communication.

        The presence of ambient signals can cause interference and hinder the accurate demodulation of tag data. To mitigate this issue, current backscatter communication systems employ strategies that restrict either signal excitation or reception. However, these approaches give rise to two main challenges. 
        \begin{itemize}
            \item i): Limited throughput: this issue affects both productive and non-productive backscatter systems. Productive systems, like FreeRider, necessitate the use of additional symbols to encode a single bit, resulting in a significant reduction in throughput. Non-productive systems, such as Interscatter, suffer from low carrier utilization, further impacting their overall throughput.
            \item ii): Dependence on single-tone exciters or extra receivers: single-tone exciters are dedicated to constant content transmission and are unable to communicate with other nodes. On the other hand, systems employing extra receivers incur higher deployment costs.
        \end{itemize}

        Bumblebee, a novel backscatter system, effectively addresses these challenges by utilizing productive BLE carriers to generate ZigBee transmissions. It achieves this by repurposing any arbitrary BLE signals for ZigBee backscattering, requiring only a single ZigBee receiver. With a carrier utilization of up to 100\%, Bumblebee offers efficient utilization of available resources. Furthermore, since smartphones, headsets, and smart speakers are already compatible with Bumblebee exciters, its deployment can be widespread.

        The paper proposing Bumblebee primarily evaluates its design using BLE version 4, which exclusively supports LE Uncoded PHY. However, the importance of LE Coded PHY, supported by BLE version 5, has grown significantly in recent years. LE Coded PHY, introduced in BLE version 5, enhances the transmission range and reliability of BLE devices by employing forward error correction (FEC) techniques. It improves data transmission robustness in challenging wireless environments and finds practical applications in IoT deployments, asset tracking systems, healthcare devices, and smart home systems. Overall, LE Coded PHY offers extended range, improved reliability, and enhanced performance in BLE-enabled applications. So in this paper, our objective is to extend Bumblebee to support LE Coded PHY in BLE version 5 and validate its effectiveness.

        Specifically, our contributions are summarized as follows:

        \begin{itemize}
            \item We extensively describe the design of Bumblebee and perform MATLAB simulations to emulate Bumblebee's behavior using BLE version 4. For comparative analysis, we also simulate Interscatter using MATLAB. Our experimental results indicate that Bumblebee achieves comparable bit error rates to Interscatter, illustrating its effectiveness without the necessity of a single-tone carrier. The primary objective of this experiment is to validate our comprehension and evaluation of Bumblebee.
            \item We analyz the properties of LE Coded PHY and found that Bumblebee can be implemented using this technology. We then conducted simulations of Bumblebee using LE Coded PHY in BLE version 5, and the results are consistent with our analysis.
        \end{itemize}

        \begin{figure}[!t]
		\centering
		\includegraphics[width=0.49\textwidth]{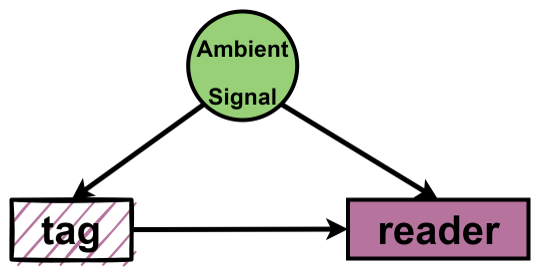}
		\caption{The architecture of ambient backscatter system.}
		\label{fig:Ambient}
	\end{figure}
	\section{BACKGROUND}
	\subsection{Ambient Backscatter System}

    Ambient backscatter is a wireless communication technique that has gained increasing attention in recent years. It enables low-power and low-cost communication for Internet-of-Things (IoT) devices by leveraging existing ambient radio signals, such as Wi-Fi, cellular, or TV signals, as the source of power and carrier for data transmission.

Unlike traditional communication systems that rely on generating dedicated signals, ambient backscatter devices utilize the surrounding ambient signals to harvest energy and reflect or modulate them to encode and transmit data. By doing so, they eliminate the need for battery-powered transmitters and significantly reduce the power consumption of IoT devices.

Ambient backscatter technology offers several advantages, including improved energy efficiency, extended battery life, and lower deployment costs. It enables seamless integration of IoT devices into existing infrastructure without the need for additional dedicated communication infrastructure. Furthermore, the ability to leverage ambient signals allows for communication in challenging environments, such as indoors or underground, where traditional wireless signals may have limited reach.

Applications of ambient backscatter can be found in various domains, including smart cities, industrial automation, environmental monitoring, and healthcare. For example, in smart cities, ambient backscatter devices can enable efficient monitoring of infrastructure and environmental parameters without requiring frequent battery replacements or extensive infrastructure upgrades.

Overall, ambient backscatter represents a promising approach to enable low-power and low-cost communication for IoT devices, offering new opportunities for connectivity in a wide range of applications. Figure~\ref{fig:Ambient} shows the architecture of ambient backscatter system.

    \subsection{Bluetooth Low
Energy}
    \label{sec:BLE}
    Bluetooth Low Energy (BLE) is a low-power version of Bluetooth technology designed to provide wireless communication capabilities for smart devices and other low-power devices. It uses a different communication protocol that can achieve battery life of up to several years, making it suitable for a variety of applications.

    \paragraph{BLE version 4 and BLE version 5.}
    The BLE 5 \cite{woolley2020bluetooth} specification has introduced three new physical layer (PHY) options to address the challenge of enhancing communication range and maximum throughput. In addition to the 1 Mbit/s Gaussian frequency shift keying (GFSK) PHY of BLE 4 (referred to as LE 1M in Bluetooth v5.0 core specification), BLE 5 includes a 2 Mbit/s GFSK PHY (called LE 2M) for high-speed transmission over short ranges, as well as two coded PHY options (known as LE Coded) with payload coding at 500 kbit/s or 125 kbit/s. The LE coded PHYs utilize GFSK modulation at a 1 Mbit/s rate, but employ a two-stage coding process: first, a forward error correction convolutional encoder is applied, and then the data is spread by the pattern mapper. Theoretically, this coding approach can enhance the link budget of coded transmissions by more than 5 dB and 12 dB compared to LE 1M, for LE coded at 500 kbit/s and 125 kbit/s, respectively. It is important to note that support for LE 1M PHY is mandatory, while the other options are optional.

    \begin{figure*}[!t] 
        \centering 
        \includegraphics[width=1\textwidth]{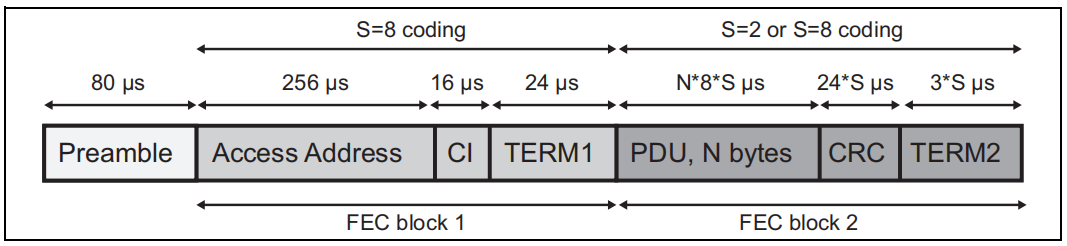}
        \caption{Link Layer packet format for the LE Coded PHY.} 
        \label{fig:PHY} 
    \end{figure*}

    \paragraph{Pachket format for the LE Coded PHY.}
    The packet format \cite{woolley2020bluetooth} shown in Figure~\ref{fig:PHY} is defined for the LE Coded PHY and is used for  packets on all physical channels. Each packet consists of the Preamble, FEC block 1, and FEC block 2. The Preamble is not coded. The FEC block 1 consists of three fields: Access Address, Coding Indicator (CI), and TERM1. These shall use the S=8 coding scheme. The CI determines which coding scheme is used for FEC block 2. The FEC block 2 consists of three fields: PDU, CRC, and TERM2. These shall use either the S=2 or S=8 coding scheme, depending on the value of the CI. The entire packet is transmitted with 1 Mbit/s modulation.

    \subsection{Zigbee}
    Zigbee is a low-power, low-data-rate, short-distance wireless communication technology mainly used for communication between IoT devices. It is characterized by low cost, low power consumption, high reliability, and easy deployment, making it suitable for home automation, industrial automation, smart health, and other fields.

    \paragraph{Commodity ZigBee Receiver.}
    ZigBee receivers utilize a quadrature demodulator for demodulating phase content. The demodulation rules are as follows: if the phase shift falls within the range [0, $\pi$], it is demodulated as '1'; if the phase shift falls within the range [-$\pi$, 0], it is demodulated as '0'.

    \section{Design of Bumblebee and simulation}
    \subsection{Design of Bumblebee}
    \label{subsec:design_bumblebee}
        \begin{figure}[!t]
		\centering
		\includegraphics[width=0.49\textwidth]{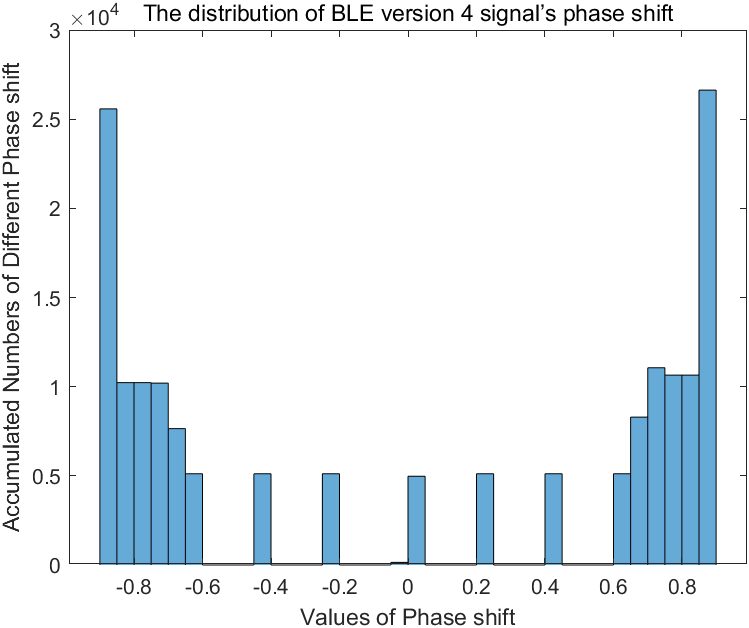}
		\caption{The distribution of BLE version 4 signal's phase shift.}
		\label{fig:ble4_dis}
	\end{figure}
    \paragraph{The phase shift of BLE version 4.}
    Its phase shift within each ZigBee chip unit is concentrated within [-1, 1]. We verify the conclusion by generating several BLE waveforms using MATLAB and analyzing their phase shifts within each ZigBee chip unit. The obtained results, depicted in Figure~\ref{fig:ble4_dis}, demonstrate that the phase shift of ambient BLE version 4 signals is predominantly concentrated within the range of [-1, 1]. This observation aligns with the findings presented in the paper that introduced Bumblebee.

    \paragraph{Data modulation.}
    Bumblebee overwrites tag data independently on ambient BLE version 4. Each phase state undergoes instantaneous shifting every Tc. Specifically, when modulating a '1', a square wave with a positive phase shift ($\Phi_0$ → ($\Phi_0$ + $\frac{\pi}{2}$)) is scheduled to control the RF switch. Conversely, when modulating a '0', a square wave with a negative phase shift ($\Phi_0$ → ($\Phi_0$ - $\frac{\pi}{2}$)) is scheduled \cite{ieee2011ieee}.

    \paragraph{Data demodulation.}
    After data modulation, the phase shift for bit '1' should be (+$\frac{\pi}{2}$ ± 1), and the phase shift for bit '0' should be (-$\frac{\pi}{2}$ ± 1). Notably, the sign of (+$\frac{\pi}{2}$ ± 1) aligns with that of +$\frac{\pi}{2}$, and the sign of (-$\frac{\pi}{2}$ ± 1) aligns with that of -$\frac{\pi}{2}$. As previously mentioned, ZigBee receivers employ a quadrature demodulator to demodulate phase content. The demodulation rules dictate that a phase shift within the range [0, $\pi$] is demodulated as '1', and a phase shift within the range [-$\pi$, 0] is demodulated as '0'. In summary, the tag data can be accurately demodulated.

    \subsection{Simulation}
    \label{exp:bumble}

        \begin{figure}[!t]
		\centering
		\includegraphics[width=0.49\textwidth]{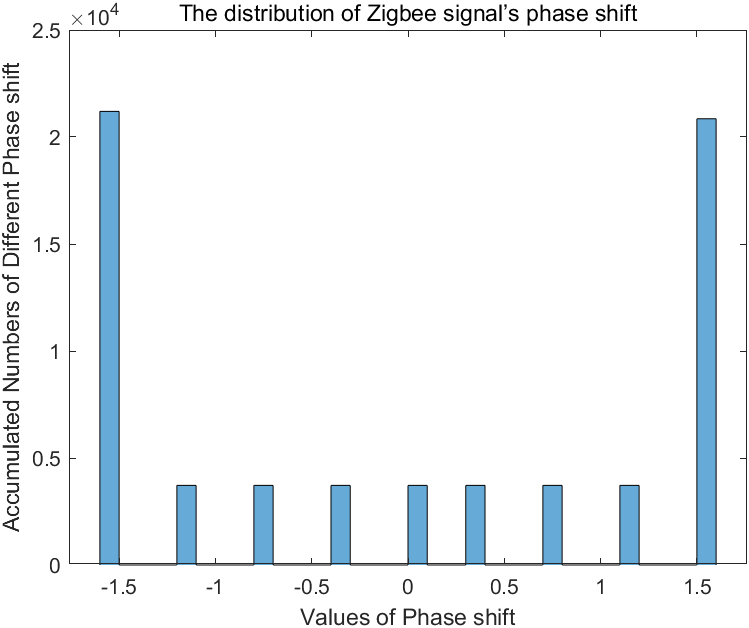}
		\caption{The distribution of Zigbee signal's phase shift.}
		\label{fig:zig_dis}
	\end{figure}
    
    \paragraph{Adjust phase shift.}
    In MATLAB simulations of Bumblebee, a crucial aspect is adjusting the phase shift of the ambient BLE signal to encode tag information. As previously discussed, when modulating a '1', a positive phase shift ($\Phi_0$ → ($\Phi_0$ + $\frac{\pi}{2}$)) needs to be applied. Conversely, when modulating a '0', a negative phase shift ($\Phi_0$ → ($\Phi_0$ - $\frac{\pi}{2}$)) should be implemented. Manually implementing this procedure can indeed be challenging.

    Fortunately, we have observed that the phase shift of the Zigbee signal distribution falls within the range of [-$\frac{\pi}{2}$, +$\frac{\pi}{2}$], as depicted in Figure~\ref{fig:zig_dis}. This aligns with the design of Bumblebee's data modulation. Consequently, we can utilize the Zigbee signal to precisely adjust the phase shift of the ambient BLE signal and effectively encode the tag information. Specifically, by adding the phase of the Zigbee signal to the phase of the ambient BLE signal, we can simulate the procedure of Bumblebee's data modulation. The core codes are shown as follows. The variable $signal$ represents the backscatter signal.
    \begin{lstlisting}[language=Matlab]
    x = phase(bleWaveform);
    y = phase(zigbeeWaveform);
    z = x + y;
    m = abs(bleWaveform);
    signal = m .* exp(1j * z);
    \end{lstlisting}

    \paragraph{Simulate Bumblebee.}
    First, we generate an ambient BLE version 4 waveform as the carrier signal. Then, using the tag information, we generate a Zigbee waveform and combine its phase with the phase of the BLE waveform. This process mimics the data modulation technique used in Bumblebee. Subsequently, we attempt to demodulate the tag information from the adjusted BLE waveform. Finally, we compare the demodulated tag information with the original tag information to calculate the bit error rate. The tag information has a length of 4KB. We repeat the experiment at various noise levels, spanning an SNR range of -15 to 15.

    \paragraph{Simulate Interscatter.}
    To simulate Interscatter, we generate a Zigbee waveform with the tag information and directly demodulate it. The length of the tag information remains 4KB. We calculate the bit error rate of Interscatter under varying noise levels within an SNR range of -15 to 15.

        \begin{figure}[!t]
		\centering
		\includegraphics[width=0.49\textwidth]{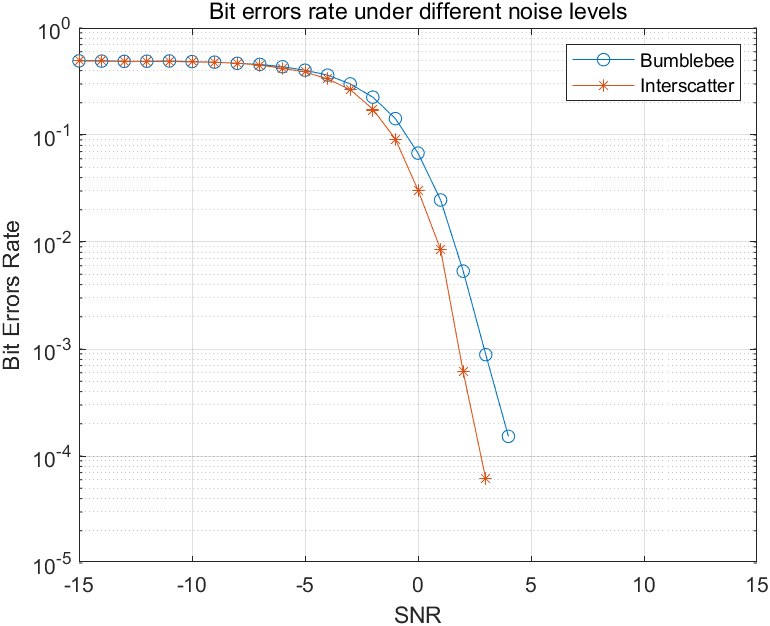}
		\caption{Experimental results of simulating Bumblebee and Interscatter.}
		\label{fig:bumble}
	\end{figure}
    \paragraph{Experimental results.}
    As depicted in Figure~\ref{fig:bumble}, both Bumblebee and Interscatter exhibit decreasing bit error rates with decreasing signal noise. For SNR values greater than 0, the bit error rate of Bumblebee remains below 0.07 and approaches 0 as the SNR decreases. This observation indicates that Bumblebee successfully demodulates the tag information from the BLE carrier, highlighting the effectiveness of its design. Furthermore, the bit error rate of Bumblebee consistently exceeds that of Interscatter, indicating that independently overwriting tag data on the ambient BLE introduces some degree of precision loss.

        \begin{figure}[!t]
		\centering
		\includegraphics[width=0.49\textwidth]{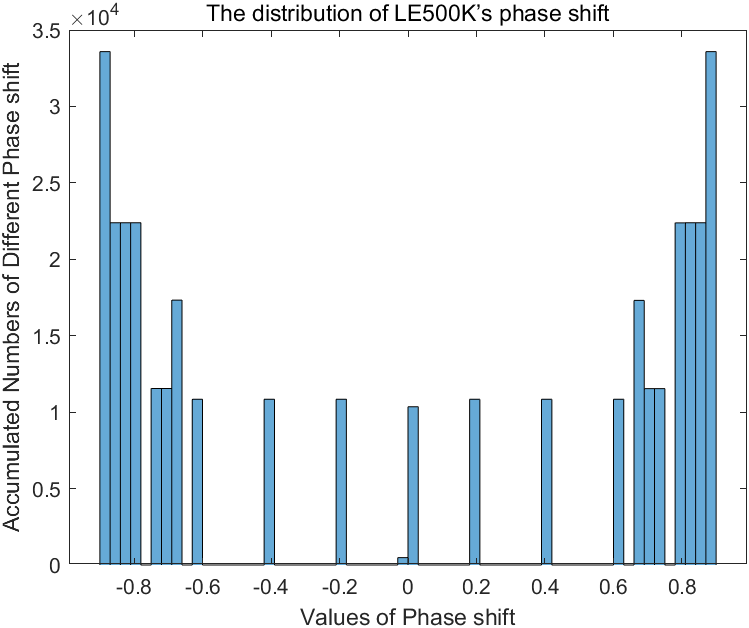}
		\caption{The distribution of LE500K's phase shift.}
		\label{fig:LE500K}
	\end{figure}

        \begin{figure}[!t]
		\centering
		\includegraphics[width=0.49\textwidth]{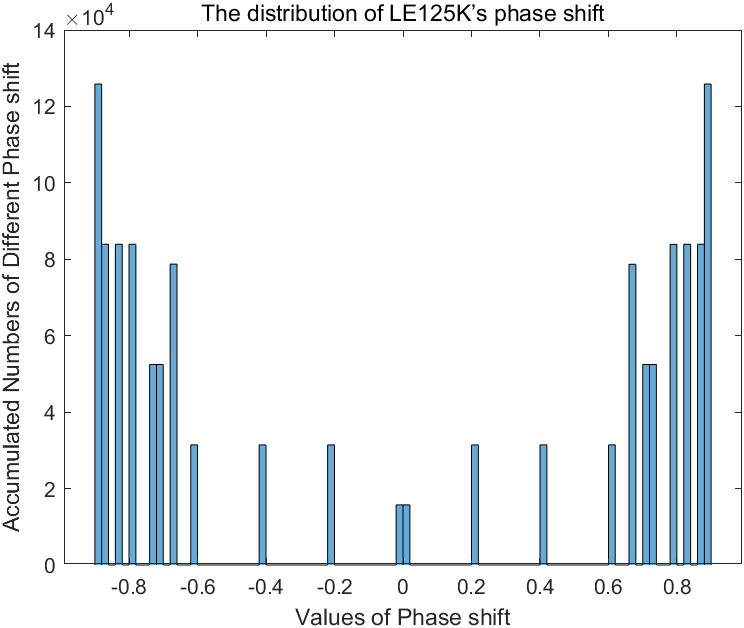}
		\caption{The distribution of LE125K's phase shift.}
		\label{fig:LE125K}
	\end{figure}

    \section{Extend Bumblebee to Support LE Coded PHY}
    \subsection{Observation and Design}
    As discussed in Section~\ref{subsec:design_bumblebee}, the suitability of a signal as a carrier for Bumblebee relies on its phase shift within each ZigBee chip unit. It is crucial for the phase shift within each ZigBee chip unit to be confined to a narrow range, allowing ample space for tag data modulation. In Section~\ref{sec:BLE}, we mentioned that the LE Coded PHY in BLE version 5 employs GFSK modulation at a rate of 1 Mbits/s, which is identical to the LE Uncoded PHY in BLE version 4. The only difference lies in the incorporation of redundant coding in LE Coded PHYs to enhance fault tolerance. Based on this, we hypothesize that LE Coded PHY and LE Uncoded PHY exhibit the same distribution of phase shifts.

    In MATLAB, the function $bleWaveformGenerator$ can be utilized to generate ambient BLE waveforms. The $Mode$ parameter of this function allows us to specify the physical mode of the BLE signal. The value "LE1M" corresponds to the LE Uncoded PHY in BLE version 4, while "LE500K" and "LE125K" correspond to two distinct LE Coded PHYs in BLE version 5. By generating waveforms using the "LE500K" and "LE125K" modes, we can observe the distribution of phase shifts associated with these PHYs.

    The phase shift distributions of "LE500K" and "LE125K" are depicted in Figure~\ref{fig:LE500K} and Figure~\ref{fig:LE125K}, respectively. It is evident that the phase shift distribution within each ZigBee chip unit for LE Coded PHYs in BLE version 5, including "LE500K" and "LE125K", falls within the range of [-1, 1]. This is consistent with the phase shift distribution of the LE Uncoded PHY in BLE version 4. Based on this observation, we can infer that LE Coded PHYs in BLE version 5 can be used as carriers in Bumblebee without requiring any modifications.

    \begin{figure}[!t]
        \centering
        \includegraphics[width=0.49\textwidth]{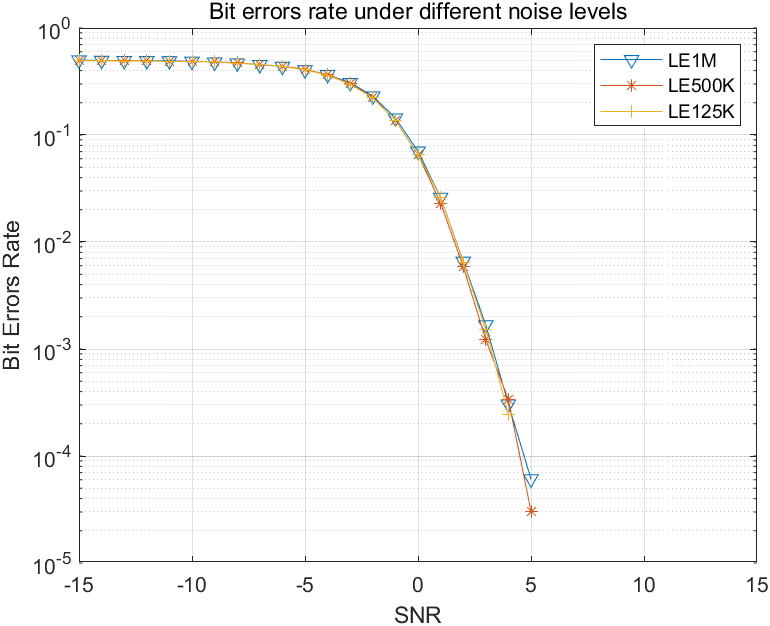}
        \caption{Experimental results of simulating Bumblebee under different physical modes.}
        \label{fig:bumblebumble}
    \end{figure}
    \subsection{Simulation}
    The simulation experiment in this section closely resembles the experiment described in \ref{exp:bumble}, with the addition of generating BLE waveforms using the parameters "LE500K" and "LE125K". The results of the experiment are presented in Figure~\ref{fig:bumblebumble}. As anticipated, the bit error rates of LE Coded PHYs in BLE version 5 align perfectly with those of the LE Uncoded PHY in BLE version 4. This confirms our initial speculation. In conclusion, Bumblebee can be deployed not only with the LE Uncoded PHY but also with the LE Coded PHY.

    \section{Future Outlook}
    \begin{itemize}
        \item In Section~\ref{sec:BLE}, we discussed the introduction of "LE2M" in BLE version 5, which utilizes LE Uncoded PHY but with GFSK modulation at a higher rate of 2 Mbit/s. This higher modulation rate may result in a different phase shift distribution compared to "LE1M", making it unsuitable as a direct carrier for Bumblebee. It is therefore worthwhile to investigate how Bumblebee can be deployed with "LE2M", allowing for its implementation with any physical mode of BLE.
        \item The experimental results depicted in Figure~\ref{fig:bumble} reveal a consistent higher bit error rate for Bumblebee compared to Interscatter. This observation suggests that the independent overwriting of tag data on the ambient BLE carrier introduces a certain level of precision loss. To enhance the performance of Bumblebee, it is crucial to investigate and identify the underlying reasons for this data loss. This understanding will enable us to improve the capabilities and effectiveness of Bumblebee accordingly.
    \end{itemize}

    \section{Conclusion}
    This paper provides a comprehensive description of the Bumblebee design and utilizes MATLAB simulations to replicate its functionality using BLE version 4. Additionally, we simulate Interscatter for comparative purposes. Our experimental findings demonstrate that Bumblebee achieves similar bit error rates to Interscatter, showcasing its effectiveness without requiring a single-tone carrier. The main objective of this experiment is to verify our understanding and evaluation of Bumblebee. Furthermore, we analyze the characteristics of LE Coded PHY and confirm its suitability for implementing Bumblebee. Subsequently, we conduct simulations of Bumblebee using LE Coded PHY in BLE version 5, and the results align with our initial analysis. Bumblebee can be deployed not only with the LE Uncoded PHY but also with the LE Coded PHY.

\bibliographystyle{ACM-Reference-Format}
\bibliography{main}
\end{document}